# Cold Dark Matter Cosmology Conflicts with Fluid Mechanics and Observations

Carl H. Gibson[1]

[1] *University of California at San Diego, La Jolla, California, 92122-0411, USA*
*Email: cgibson@ucsd.edu*

**ABSTRACT**

Cold dark matter hierarchical clustering (CDMHC) cosmology based on the Jeans 1902 criterion for gravitational instability gives predictions about the early universe contrary to fluid mechanics and observations. Jeans neglected viscosity, diffusivity, and turbulence: factors that determine gravitational structure formation and contradict small structures (CDM halos) forming from non-baryonic dark matter particle candidates. From hydro-gravitational-dynamics (HGD) cosmology, viscous-gravitational fragmentation produced supercluster ($10^{46}$ kg), cluster, and galaxy-mass ($10^{42}$ kg) clouds in the primordial plasma with the large fossil density turbulence ($\rho_0 = 3 \times 10^{-17}$ kg m$^{-3}$) of the first fragmentation at $10^{12}$ s, and a protogalaxy linear and spiral clump morphology reflecting maximum stretching near vortex lines of the plasma turbulence at the $10^{13}$ s plasma-gas transition. Gas protogalaxies fragmented into proto-globular-star-cluster mass ($10^{36}$ kg) clumps of protoplanet gas clouds that are now frozen as earth-mass ($10^{24-25}$ kg) Jovian planets of the baryonic dark matter, about 30,000,000 rogue planets per star. Observations contradict the CDMHCC prediction of large explosive Population III first stars at $10^{16}$ s, but support the immediate gentle formation of small Population II first stars in globular-star-clusters from HGD.

## 1. INTRODUCTION

The Jeans (1902) acoustic criterion for self-gravitational structure formation is derived by linear perturbation stability analysis of the Euler fluid momentum equations with gravity, giving linear acoustic equations by neglecting diffusion, viscous forces and turbulence. Density $\rho$ is set to zero (the Jeans swindle) to derive a critical (Jeans) length scale $L_J = V_s / (\rho G)^{1/2}$, where $V_s$ is sound speed and G is Newton's gravitational constant, interpreted by Jeans and astrophysicists (e.g. Springel et al. 2005) as the only gravitational stability criterion forbidding gravitational structures at scales smaller than $L_J$. Sir James Jeans (1929) claimed observational support for his theory is provided by spiral galaxies. He assumed hot gas from a (mysterious) central spinning source collapses to stars in such galaxies only when chilled by the coldness of space, thus reducing $L_J$.

This evidence weakened as better telescopes showed the "stars" observed were globular star clusters (GCs) containing millions of stars. Observational and theoretical support for a central gas source in spiral galaxies failed to materialize. Modern fluid mechanics requires additional gravitational structure formation criteria. When viscosity, diffusion and turbulence are included in the analysis, a hydro-gravitational-dynamics (HGD) cosmology emerges that is in better agreement with recent observations of the early universe than CDM scenarios that assume collisionless fluids (Gibson 1996, 2000, 2004, 2005, 2006).

From HGD cosmology (Figure 1), proto-galaxies and proto-galaxy-clusters formed by gravitational fragmentation of plasma at the Schwarz viscous scale $L_{SV} = (\gamma \nu / \rho G)^{1/2}$, where $\gamma$ is the rate-of-strain, $\nu$ is the kinematic viscosity, $\rho$ is the density, and G is Newton's constant, preserving various parameters of these irreversible transitions as fluid mechanical fossils.

## 2. DILEMMAS OF JEANS' THEORY

The Jeans criterion presents major dilemmas for hot big bang cosmology in an expanding universe. In the H-He plasma epoch, the Jeans scale of the plasma $L_J$ always exceeds the "horizon" scale of causal connection $L_H = ct$ because the enormous sound speed $V_s = c/3^{1/2}$ is of order the speed of light $c$, where $t$ is the time, so plasma structures are forbidden. The primordial gas at transition has a Jeans mass of a million stars ($M_{J_o} = 10^{36}$ kg), $t = 10^{13}$ s (300,000 years), and $T_0 = 4000$ K.

The Jeans mass $M_J = M_{J_o}(T/T_0)^{3/2}\rho/\rho_0$ from the Jeans criterion, so the first stars are large and explosive, Madau (2006), and can form only after a "dark ages" period of about 300 Myr ($10^{16}$ s) to permit sufficient cooling and density decrease. These Population III superstars rapidly disintegrate as dusty class II supernovas to re-ionize the gas and explain its absence in absorption spectra from early quasars, Mori and Umemura (2006). However, recent γ-ray spectra from blazars refute the intense high-energy starlight expected, Aharonian et al. (2006). The intergalactic medium is free of any high density of Pop-III photons that would scatter blazar γ-rays by pair production. Chemical searches also fail to detect the large amounts of stardust Pop-III superstars would have produced had they existed.

### 3. HGD COSMOLOGY

After the big bang and a fragmentational cascade to form protogalaxies (Fig. 1, Gibson 2006), HGD cosmology predicts primordial gas planets were the first collapsing gravitational objects. Superstars of CDM cosmology are not necessary to explain the missing gas since it was sequestered as planets. Some of the planets gently formed larger planets and small stars in dark proto-globular-clusters (PGCs) of dense, dark, cooling proto-galaxies (PGs). Freezing weakly collisional PGC clumps expanded by diffusion from PGs to form >$10^{21}$ m galaxy halos of baryonic dark matter (BDM), with diffusivity $D \sim 10^{27}$ m$^2$ s$^{-1}$ calculated from $L_{SD}$ and the size and ρ of the halos.

Flat rotation curves observed outside the $10^{20}$ m spiral galaxy cores reflect this huge diffusivity. Luminous >$10^{20}$ m accretion disks and spiral patterns of star formation in globular-star-clusters (GCs) reflect torques developed in the chain-like morphology of the proto-galaxy-clusters as they split into spiral galaxies from the expansion of space. The tiny first stars still exist in GCs and dim proto-GCs that are bound by gravity and friction from the evaporated gas provided by heating of their trillion frozen gas planets.

### 4. CDM COSMOLOGY

Orthodox collisionless cosmology explains galaxy and galaxy cluster formation using "cold dark matter" (CDM): a postulated non-baryonic fluid of collisionless massive particles produced during the nucleosynthesis epoch with particle velocity $v_{CDM} < c$. Based on Jeans theory and collisionless N-body simulations, it is assumed *deus ex machina*, Binney & Tremaine (1987), that gravitationally bound CDM chunks form and merge gravitationally like point mass N-bodies are forced to do in computer simulations. Large mergers ("halos") of CDM chunks then serve as growing baryon bottles that guide the gravitational formation of larger structures by hierarchical clustering (CDMHC) for 300 Myr dark ages without stars and billions of years without galaxies or galaxy clusters.

Pop-III superstars form as the gas cools by radiation and promptly explode in the merging CDM-halos, Mori et al. (2004). Jupiter mass gas planets are ruled out by the Jeans criterion because they require an impossible cooling of their H-He gas to 0.004 K to permit Jeans instability. However, Jovian planets circling the sun in outer orbits and now detected in great multitudes in tight orbits around other stars, Lovis et al. (2006), contradict the Jeans acoustic criterion and support the basic tenet of HGD that all the hot gas after transition fragments into dense PGC globular star cluster mass clouds of Earth mass PFP primordial fog particles.

How can the postulated CDM chunks remain gravitationally bound and merge to form larger chunks? Because fundamental particles of CDM are assumed to be collisionless and the chunks are not point masses but must occupy finite volumes, CDM chunks have no physical mechanisms to stick to each other on collision or resist tidal forces $F_{TIDAL} = d_{CDM} G m_{CDM} M_{CDM}/r^3$, where $d_{CDM}$ is the diameter of CDM chunks of mass $m_{CDM}$ attracted by a larger chunk of mass $M_{CDM}$ at distance r.

The tidal distortion $x_{TIDAL}$ per free fall oscillation is $x_{TIDAL}/d_{CDM} = (t/\tau)^2 (D_{CDM}/r)^3 M_{CDM}/2m_{CDM} > 1$. Small $m_{CDM} \ll M_{CDM}$ chunks are torn to smaller and smaller masses $m_{CDM} \to m_{CDM-particle}$ by any larger masses $M_{CDM}$ in a few free fall times $\tau = (\rho G)^{-1/2}$ (Figure 2), contradicting the hierarchical merging process postulated in CDM cosmology. Chunks of CDM of any size will not remain gravitationally bound but will expand indefinitely by diffusion. The CDMHC hypothesis is therefore false. Observations show no galaxies with cusp-like central concentrations of density reflecting the collisionless core collapse expected for CDM halos of CDM chunks.

Gravitationally bound CDM chunks are claimed (Binney & Tremaine 1987, chapter 4) to have large stability periods $\tau_{CDM} = \tau N/\ln N$ much longer than $\tau$ if N is large, where N is the number of CDM particles. The claim is unwarranted. Any collection of collisionless particles with any velocities will grow in size indefinitely by diffusion. Assume a spherical CDM-chunk initially with perfectly cold, motionless, particles. All particles fall toward the center of the sphere arriving at time $t \approx \tau$ as the density of the shrinking sphere rapidly increases, the collisionless assumption fails, the particles exchange momentum and energy, and diffusion increases the size of the chunk to a scale $L_{SD} = (D^2 / \rho G)^{1/4}$ larger than its initial size (Figure 3). The free fall time $\tau$ increases as the CDM-chunk size increases.

Galaxies of stars in apparent virial equilibrium are claimed, Binney & Tremaine (1987), as prime examples of permanent ($\tau_{galaxy} \sim \tau N / \ln N$) gravitationally-bound collisionless systems. However, stars are actually not collisionless because on average 30 million planets surround each star as the interstellar medium from which stars are formed. The planets supply gas and friction in response to any rapid relative motions, and reveal themselves as contrails of stars (as in the Tadpole, Mice, and Antenna systems) formed in galaxy baryonic dark matter halos when galaxies frictionally merge, Gibson & Schild (2003), Gibson (2006).

Schild (1996) has shown by observations of quasar microlensing twinkling frequencies that the lensing galaxy mass is dominated not by stars but by planets, which he proposed as the galaxy missing mass. Numerous claims that the collisionless tidal tails of computer simulations of galaxy mergers have been observed in nature are therefore false. Star trails between merging and separating galaxies are never collisionless tidal tails but evidence of galaxies triggering star formation as they move through each other's highly combustible BDM halos.

Diffusivity $D = Lv$ in a gas is mean free path $L_{MFP} = m_{particle} / \rho_{gas} \sigma_{collision}$ times the particle velocity $v_{particle}$, so whatever fundamental particle emerges as that of the unknown non-baryonic dark matter (NBDM) will certainly have a very large $L_{MFP}$ and large diffusivity $D_{NBDM}$ since the collision cross sectional area $\sigma_{NBDM}$ is small ($< 10^{-40} m^2$) and the particle velocity $v_{NBDM} \sim c$ large.

From the size and density of galaxy clusters and the Schwarz diffusive scale $L_{SD} = (D^2 / \rho G)^{1/4}$, the NBDM particle mass appears to be about $10^{-35}$ kg, Gibson (2000), with diffusivity $D_{NBDM} \approx 10^{30}$ m$^2$ s$^{-1}$ and a total NBDM mass about 30 times that of all baryons. With these values, the NBDM diffuses to $L_{SD}$ scales somewhat larger than the horizon scale $ct$ during the plasma epoch and begins fragmentation at galaxy cluster halo scales of about $10^{23}$ m only after the transition to gas. In the hot plasma period between $10^2$ s and $10^{11}$ s when energy dominated mass it is likely that the NBDM particles coupled to the plasma to dominate momentum transfer in both fluids, preventing both turbulence and gravitational structure formation. A mechanism like the high temperature MSW scattering of neutrinos by electrons is anticipated, Mikheyev & Smirnov (1985), Wolfenstein (1978).

No cosmological dilemmas arise if collisional, nonlinear fluid mechanics is applied to cosmology rather than the unrealistic simplifications of Jeans 1902 and CDMHC. Self-gravitating fluids are absolutely unstable at density minima and maxima and form structures unless prevented by fluid forces or diffusion. Fossil turbulence[*] from inflation preserves patterns of the big bang turbulence, Gibson (2004).

Only viscous and turbulence forces were relevant in the primordial plasma until buoyancy forces of the first self-gravitational structures (at $10^{12}$ s) appeared to inhibit large scale turbulence[†] and to preserve the plasma density ($\rho_0 \sim 3 \times 10^{-17}$ kg m$^{-3}$) and rate-of-strain ($\gamma_0 \sim 10^{-12}$ rad s$^{-1}$) as fossils of the turbulence. Diffusion dominates gravity for the NBDM in the plasma epoch since $L_{SD} > L_H = ct$ during this period.

Coupling to the plasma ceased soon after $10^{11}$ s. Condensation on density maxima was inhibited by the rapid expansion rate of the early universe, so the first gravitational structures were formed by fragmentation at density minima when the viscous Schwarz scale $L_{SV} = (\gamma v / \rho G)^{1/2} > L_H$ decreased to less than $L_H = ct$ at $10^{12}$ s, forming voids and clouds of plasma when viscous forces matched those of gravity, where $\gamma_0 \sim 1 / t_0$ is the rate of strain ($10^{-12}$ s$^{-1}$) and $v$ is the kinematic viscosity ($4 \times 10^{26}$ m$^2$ s$^{-1}$).

Substituting these values gives $L_{SV} \approx L_H = 3 \times 10^{20}$ m at $t_0 = 10^{12}$ s with plasma mass $\sim 10^{45}$ kg corresponding to that of a galaxy supercluster. The horizon scale

---

[*] Fossil turbulence is defined as a fluctuation in any hydrophysical field produced by turbulence that persists after the fluid ceases to be turbulent at the scale of the perturbation.

[†] Turbulence is defined as an eddy-like state of fluid motion where the inertial vortex forces of the eddies $\vec{v} \times \vec{\omega}$ are larger than any other forces that tend to damp the eddies, $\vec{v}$ is the velocity and $\vec{\omega}$ is the vorticity. Turbulence by this definition always cascades from small scales to large.

Reynolds number $Re_H = c^2 t / \nu$ ~250 for the plasma was only slightly above critical, so the turbulent and viscous Schwarz scales $L_{ST} = \varepsilon^{1/2} / (\rho G)^{3/4} \approx L_{SV}$ were nearly identical. The viscous dissipation rate $\varepsilon$ was ~800 m² s⁻³. Patterns of big bang fossil turbulence that triggered the fragmentation are preserved by the cosmic microwave background, Gibson (2005).

## 5. OBSERVATIONS

Gravitational fragmentation of the plasma continued to smaller scales, forming protogalaxy mass clouds ($10^{42}$ kg) prior to gas formation, with density $\rho_0$ and an initial protogalaxy size $3 \times 10^{19}$ m (1 kpc). Density minima and maximum rates of strain $\gamma$ provide sites for fragmentation near vortex lines of the weak turbulence, giving dim straight chains of $6 \times 10^{19}$ m protogalaxy clumps (Figure 4) and spiral clusters of protogalaxy clumps as observed by the Hubble Ultra Deep Field, Elemegreen et al. (2005).

Direct numerical simulations of turbulence, Nomura and Post (1998), show maximum $\gamma$ values required for gravitational structure formation by fragmentation occur near turbulence vortex lines (Figure 5). The pressure Hessian $\Pi_{ij} = \partial^2 p / \partial x_i \partial x_j$ describes nonlocal interactions of the vorticity and the rate of strain tensor, resulting in maximum values of $(\gamma_{ij})_{eigen}$ along the vortex lines $(\gamma_{ij})_{eigen} = (+, -, -)$ but with the average dominated by stretching in two directions near the ends $(\gamma_{ij})_{eigen} = (+, +, -)$. The resulting protogalaxy morphology to be expected from HGD is shown in Fig. 5.

The Schwarz viscous scale of the 4000 K hot gas at $10^{13}$ s was about $10^{14}$ m giving a small-planet mass of $10^{24-25}$ kg. Simultaneous fragmentations at $10^{36}$ kg were from radiative heat transfer density changes at the Jeans scale $L_J$, but not by the Jeans 1902 mechanism. The viscous dissipation rate $\varepsilon$ was a very gentle ~$10^{-12}$ m² s⁻³.

Gravitational condensation of matter first began in the universe as planetary cloud collapse with $\tau_0 = (\rho_0 G)^{-1/2}$ in proto-globular-clusters (PGCs). The collapse time $\tau_0$ was $3 \times 10^{13}$ s (600,000 years) which is also the time to first star formation in the tidally agitated PGC cores near protogalaxy centers, not the 300 million dark ages years ($10^{16}$ s) predicted by CDMHC after assembly of CDM halos and cooling of the gas to permit the first stars as Population III superstars by the Jeans criterion.

## 6. CONCLUSIONS

The standard cosmology based on collisionless fluid mechanics and cold dark matter is obsolete and should be abandoned because it gives predictions contrary to observations. Hydro-gravitational-dynamics cosmology predictions give protosupercluster to protogalaxy formation in the plasma epoch ($10^{12}$ to $10^{13}$ s) by fragmentation triggered by density patterns of big bang turbulent mixing preserved as the first fossils of turbulence.

Stars form from planets that fragmented in $10^{36}$ kg clumps immediately after the plasma to gas transition. The high $10^{12}$ s baryonic density is preserved as that of globular star clusters. Gentle motions prevailing at the $10^{13}$ s time of first stars made possible their small sizes. The interstellar medium and the baryonic dark matter are proto-globular-star-clusters of frozen small planets, as predicted by HGD cosmology and observed by quasar microlensing.

CDM hierarchical clustering cosmology gives gravitationally bound structures forming first at small scales and clustering by impossible fluid mechanics. These structures gradually form stars that are much too large and too late, and that explode to form light that is found not to exist. Galaxies and galaxy clusters are observed at times earlier than predicted by CDMHC cosmology, but as expected from HGD cosmology.

## 7. FIGURES

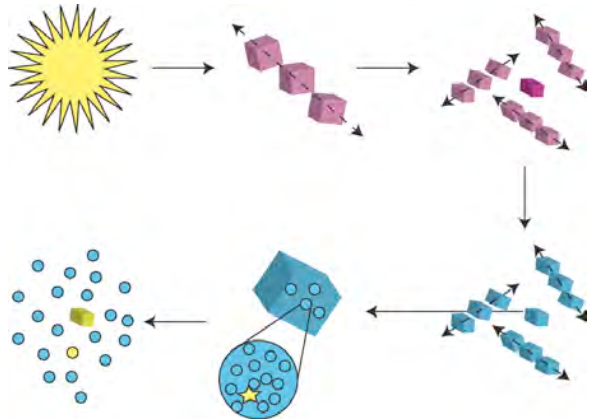

Figure 1. Gravitational structure formation according to HGD cosmology (Gibson 2006). Proto-supercluster-voids form at minimum density points along vortex lines of weak turbulence in the hot plasma at $10^{12}$ s (top center). Fragmentation continues to $10^{13}$ s when plasma proto-galaxies turn to gas (right top bottom). These fragment into proto-globular-star-cluster (PGC) clouds of primordial-fog-particle planets (PFPs) that form the first stars (bottom center). The freezing PGCs diffuse out of the galaxy core to form its baryonic dark matter halo (bottom left).

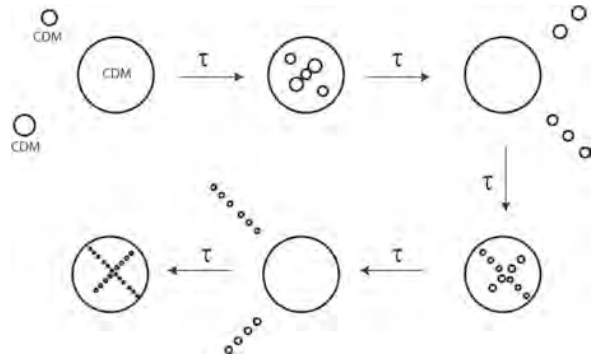

Figure 2. Three hypothetical CDM chunks interact gravitationally. Because they have no sticking mechanism, the collisionless CDM chunks are shredded to atoms by tidal forces and cannot grow by hierarchical clustering.

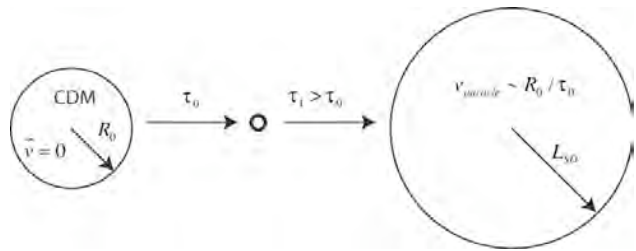

Figure 3. A perfectly cold chunk of CDM with motionless particles collapses in a free fall time $\tau_0$ to a size small enough for the collisionless assumption to fail. The "chunk" then expands to size $L_{SD}$ with particle kinetic energies determined by the initial gravitational potential energy.

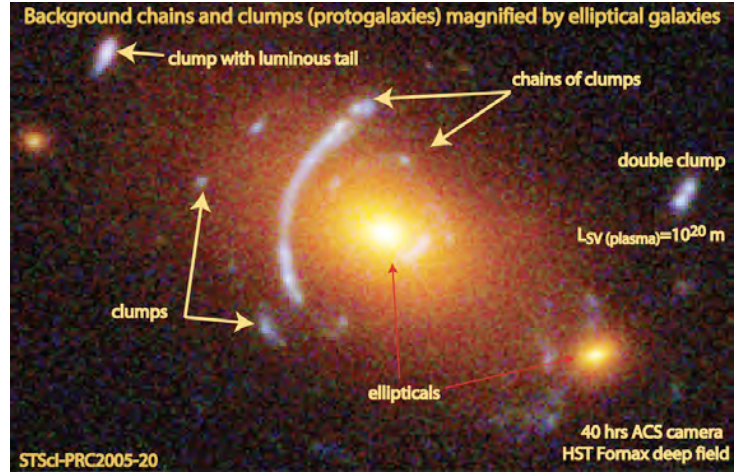

Figure 4. Elliptical galaxies magnify light from the first galaxies, which are in linear structures reflecting viscous-gravitational formation along vortex lines of the weakly turbulent plasma just before transition to gas (Fig. 1). Star formation is triggered in the BDM between the 2 kpc ($6 \times 10^{19}$ m) protogalaxies (clumps).

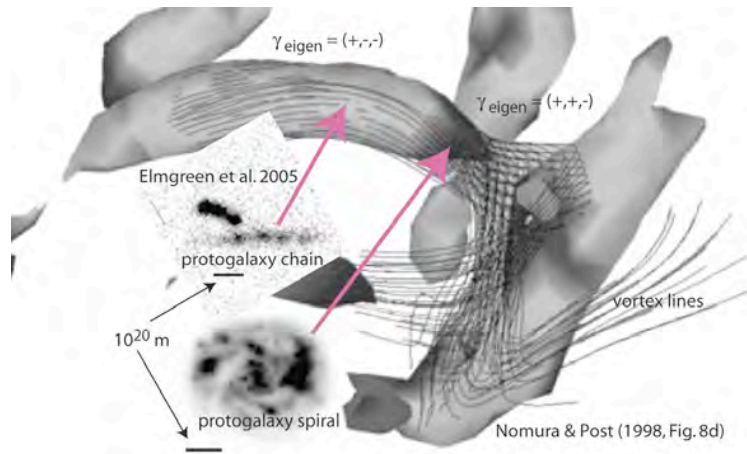

Figure 5. Direct numerical simulations of turbulence by Nomura & Post (1998) show maximum stretching occurs on vortex lines where $\left( \gamma_{ij} \right)_{eigen} = (+,-,-)$, but the average stretching comes from the end of vortex tubes where $\left( \gamma_{ij} \right)_{eigen} = (+,+,-)$. From HGD, protogalaxy formation in the turbulent plasma is triggered where the least principal rate of strain (the stretching rate) is large, which can explain the clump chains and clump spirals observed by Elmgreen et al. (2005). The scale $10^{20}$ m is that of central bulge regions of galaxies, suggesting these are fossils of their protogalaxy clump origins.